\documentclass[aps,twocolumn,showpacs,preprintnumbers,superscriptaddress]{revtex4-1}

\usepackage{array}
\newcolumntype{L}[1]{>{\raggedright\let\newline\\\arraybackslash\hspace{0pt}}m{#1}}
\newcolumntype{C}[1]{>{\centering\let\newline\\\arraybackslash\hspace{0pt}}m{#1}}
\newcolumntype{R}[1]{>{\raggedleft\let\newline\\\arraybackslash\hspace{0pt}}m{#1}}

\usepackage{amsmath}
\usepackage{amssymb}
\usepackage{amstext}
\usepackage{amsopn}
\usepackage{amsfonts}
\usepackage{amsxtra}
\usepackage[colorlinks]{hyperref}
\usepackage{url}
\usepackage{mathrsfs}
\usepackage[dvips]{graphicx}
\usepackage{relsize}
\usepackage{mathtools}
\usepackage{float}
\usepackage{bm}
\usepackage[dvipsnames]{color}
\numberwithin{equation}{section}

\newcommand{\mb}{\mathbf}

\newcommand{\h}{\hbar}

\newcommand{\ee}{\varepsilon}
\newcommand{\kk}{\mathbf{k}}

\newcommand{\la}{\langle}
\newcommand{\ra}{\rangle}

\newcommand{\e}{\varepsilon}
\newcommand{\s}{\sigma}

\newcommand{\om}{\omega}
\newcommand{\Om}{\Omega}
\newcommand{\al}{\alpha}

\newcommand{\de}{\delta}
\newcommand{\ka}{\kappa}
\newcommand{\D}{\Delta}

\newcommand{\E}{\mathscr{E}}

\def\mathclap#1{\text{\hbox to 0pt{\hss$\mathsurround=0pt#1$\hss}}}

\begin{document}
\def \brho{{\hbox{\boldmath $\rho$}}}
\def \beps{{\hbox{\boldmath $\epsilon$}}}
\def \bdelta{{\hbox{\boldmath $\delta$}}}

\title{Carrier concentrations and optical conductivity of a band-inverted semimetal in two and three dimensions}
\author{Zoran Rukelj }
\email[]{zrukelj@phy.hr}
\affiliation{Department of Physics, University of Fribourg, 1700 Fribourg, Switzerland}
\affiliation{Department of Physics, Faculty of Science, University of Zagreb, Bijeni\v{c}ka 32, HR-10000 Zagreb, Croatia}
\author{Ana Akrap}
\email[]{ana.akrap@unifr.ch}
\affiliation{Department of Physics, University of Fribourg, 1700 Fribourg, Switzerland}
\date{\today}

\begin{abstract}

Here we study the single-particle, electronic
transport and optical properties of a gapped system described by a simple two-band Hamiltonian with inverted valence  bands. 
We analyze its properties in the three-dimensional (3D) and the two-dimensional (2D) case.
The insulating phase changes into a  metallic phase when the band gap is set to zero. The metallic phase in the 3D case is characterized by a nodal surface. This nodal surface is equivalent to a nodal ring in two dimensions.
Within a simple theoretical framework, we calculate the density of state, the total and effective charge carrier concentration, the Hall concentration and the Hall coefficient, for both 2D and 3D cases.
The main result is that the three concentrations always differ from one another in the present model.
These concentrations can then be used to resolve the nature of the electronic ground state.
Similarly, the optical conductivity is calculated and discussed for the insulating phase. 
We show that there are no optical excitations in the metallic phase.
Finally, we compare the calculated optical conductivity  with the rule-of-thumb derivation using the joint density of states.
\end{abstract}

\maketitle

\section{Introduction}

The topological quest within the solid state physics is to identify properties that originate from the so called non trivial topology of the Bloch bands \cite{RevModPhys.82.3045, RevModPhys.82.1959}.
Many systems have been explored. Most familiar are the Dirac and Weyl semimetals, which contain a whole array of candidate materials \cite{Armitage2018a}.
Their common underlying feature is that the valence and conduction bands touch in one or more isolated points in the Brillouin zone
\cite{Polatkan2020, Habibi2021, PhysRevB.95.161112}. 
A natural extension of the band point-touch is a band line-touch. Such materials are known as nodal-line semimetals \cite{Carbotte2016, Rhim2016, Shao2019}. 
Analogously, extending the idea of the nodal line semimetal leads to the nodal surface semimetal (NSSM), in which the bands touch over a surface spanned in the Brillouin zone \cite{Wang2018a}. 
This surface is equivalent to a line in the 2D case \cite{Jin2020}. 

In this article, we address the simplest case of NSSM in 3D and 2D. We also explore their gapped phase, which we refer to as the gapped semimetal phase (GSM).
We analyze the single-particle intraband and interband properties of the system described by a two-band Hamiltonian. 
The Hamiltonian contains three free parameters and it describes the gapped (GSM) and the metallic (NSSM) phase. The energy bands' main feature is the inverted shape up to some critical energy \cite{McCann2006, Guinea2006} and a parabolic free-electron-like dispersion at energies beyond the band inversion.
Our intention is twofold.
First, we want to answer the question: Can the experiments such as the electronic transport and optical measurements resolve the two possible ground states, gapped (GSM) and semimetallic (NSSM)?
This tackles the main problem inherent to nearly all topological materials. Their intrinsic energy scales---energy intervals within the Bloch bands---where their 
topological properties can be observed is small, often not more than several milli-electron-volts.
This makes it challenging to experimentally distinguish between different possible ground states. 

Second, we want to show how the DC and optical properties differ in the 3D and 2D cases of such semimetals.
The DC quantities include the density of state per unit volume, the total and the effective concentration of the charge carriers,  the Hall concentration and the Hall coefficient.
We  show that very generally those concentrations differ for both GSM and NSSM phases, in 2D and in 3D.  
Therefore, by comparing the Drude weight and  the Hall concentration with the total concentration, we can determine the nature of the ground state.
The discrepancies between the GSM and NSSM phases are even more evident in the optical conductivity. 
The NSSM phase has no optical excitations since the amplitude of the interband current matrix element is proportional to the 
energy gap $\D$.
In the GSM case, we calculate the  real part of the optical conductivity within the vanishing and finite interband relaxation constant approximation. 

Without interband relaxation, the optical conductivity has the same shape for 3D and 2D when incident photon energy  is just above the band gap. 
It has a square-root singularity $ {\rm{Re}} \, \s(\Om) \propto (\Om-\D)^{-1/2}$. For large energies, the conductivity scales with the dimension $D$ as 
$ {\rm{Re}} \, \s(\Om) \propto \Om^{D/2 -3}$.

Knowing the specific Bloch momentum dependence of the interband current vertex, we show when one can  use the joint density of state rule-of-thumb calculation \cite{gruner} to estimate the real part of the optical conductivity. 

This paper is structured as follows: In the first part of the article we specify the model Hamiltonian for the GSM and NSSM phase. We then define and calculate for the 3D case the density of states, the  three concentrations of charge carriers (total, effective and Hall) and the real part of the optical conductivity. 
Finally, we calculate all these quantities for the 2D case. 
\section{Two-band Hamiltonian}\label{sec1} 
As a starting point, we define a continuum $2\times 2$ isotropic Hamiltonian matrix that describes the general form of GSM and consequently the NSSM phase \cite{Pal2016, Zhang2016, Carrington2011}. 
The Hamiltonian is
\begin{equation} \label{ham1}
  \hat{H} = (A- B k^2)\s_z + C\s_x,
\end{equation}
where $\s_x$ and $\s_z$ are the Pauli matrices  and 
$A$ and $C$ are positive constants representing the gap parameters. The $A- B k^2$ is the ``inverted part''. It is the simplest isotropic form of a 
nodal surface, with $k^2$ being the square of the total Bloch wave vector.
The positive parameter $B$ can be written in a more familiar way  as $B = \h^2/(2m^*)$. The effective mass $m^*$ will come in handy when the DC properties are discussed in the next section.
The  Hamiltonian (\ref{ham1}) is invariant under spatial inversion, and since it is a real matrix it is also invariant under time reversal. The latter also implies vanishing Berry 
curvature \cite{Gradhand2012, Sinitsyn2007} making this system topologically trivial.
This Hamiltonian is a simplified variant of the Bernevig--Hughes--Zhang Hamiltonian \cite{BernevigScience06, Peres2013},
 with a constant electron-hole coupling described by the parameter $C$. 

The diagonalization of Eq.~(\ref{ham1}) is straightforward. It gives an electron-hole symmetric eigenvalues
\begin{equation}\label{ham3}
  \ee^{c,v}_{\kk} = \pm  \sqrt{( A-  B k^2)^2 +  C^2  }.
\end{equation}
The indices $c$ and $v$ stand for conduction band (plus sign) and valence band (minus sign), respectively.

To make the analysis of the electron properties originating from Eq.~(\ref{ham1}) as general as possible, we  scale the eigenvalues Eq.~(\ref{ham3}) to the gap parameter
$A$ and introduce dimensionless
substitutions. These substitutions are a dimensionless gap $\D= C/A $ and a dimensionless wave vector $\ka^2 = k^2 B/A$.
In this way, the eigenvalues above become much simpler:
\begin{equation}\label{ham4}
  \om_\ka = \pm  \sqrt{( 1 - \ka^2)^2 +  \D^2 },
\end{equation}
with the definition  $\om_\ka= \ee^{c,v}_{ \kk}/ A$. 

\begin{figure}[tt]
\includegraphics[width=.48\textwidth]{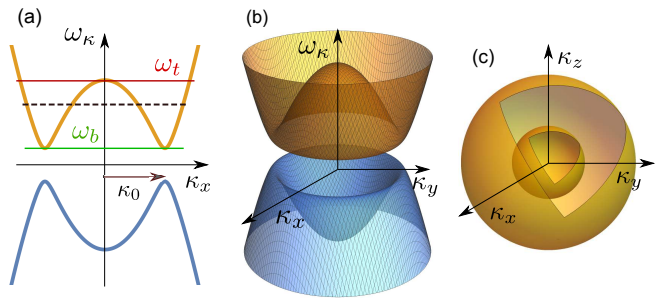} 
\caption{ (a) Valence bands Eq.~(\ref{ham4}) of the GSM case in 1D. The inverted conduction band (orange) spans the energy range between $\om_b$ (green) and $\om_t$ (red). The dashed line indicates the Fermi energy $\om_F$.
In the NSSM phase, the bands touch on the surface of a circle of radius  $\ka_0$, while in the GSM phase $\ka_0$ is the position of minimum (maximum) of the conduction (valence) band. 
(b)  Analogous valence bands in 2D. 
(c) The Fermi surface of the 3D system in the case of a partially filled conduction band with the Fermi energy in the range $\om_b < \om_F < \om_t$ as depicted by the black level 
in figure (a).}
\label{f1}
\end{figure}

The dispersions given by Eq.~(\ref{ham4}) are shown in  Fig.~\ref{f1}. Bands for the 1D $\kk = (k_x,0,0)$ and 2D case are shown, together with the Fermi surface in 3D. The height of the band inversion is described by the parameter $A$, while $C$ gives the minimum band separation. In dimensionless units this differentiates the two phases, GSM and NSSM. From  Eq.~(\ref{ham4}) for GSM we have $\D >0$, and for NSSM $\D =0$.
In the three-dimensional NSSM case the two bands touch along the spherical surface of radius $\ka_0 = 1$. The sphere becomes a circle in 2D. In 1D case we can depict
$2\ka_0$ in Fig.~\ref{f1}a, as the distance between the two points where the bands touch.

If the value of the Fermi energy $\om_F$ is restricted to  $\om_b < \om_F <\om_t$, the Fermi surface consists of two concentric spheres in 3D or concentric circles in 2D. 
With 
\begin{equation}\label{ham5}
\om_b = \D, \hspace{5mm} \om_t = \sqrt{1+\D^2},
\end{equation}
we designate the energy belonging to the bottom ($b$) and the top ($t$) of the inverted band as shown in Fig.~\ref{f1}(a) and (b). Energies (\ref{ham5}) to a great extent determine  the specific behavior of the DC concentrations and the optical response, as it will be shown in the following sections.
These energies, $\om_b$ and $\om_t$, define the energy interval where the inverted bands directly influence the electronic transport.

\section{3D case}\label{sec2} 

\subsection{Density of states}\label{dos}

\begin{figure}[tt]
\includegraphics[width=.45\textwidth]{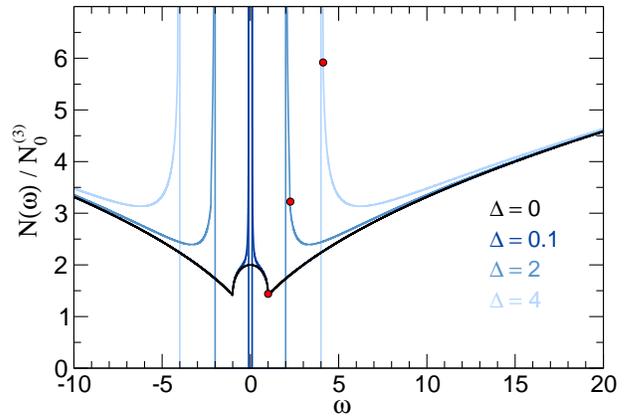} 
\caption{The density of state Eq.~(\ref{g6}) of the 3D system derived from the energy dispersion Eq.~(\ref{ham3}). DOS is a function of a dimensionless parameter $\om$ and is 
plotted for several values of the gap parameter, $ \om_b = \D$. For the case of $\D=0$ (NSSM), the dome in DOS is clearly visible between the energies $(-1,1)$ (black line) with a maximum height of $2N_0^{(3)} $. For $\D>0$ at the energy $\om = \om_b$, DOS has a square-root singularity, and the value at $\om=\om_t$ is given by red circles. For energies $\om \gg \om_t$, DOS is $\propto \sqrt{\om}$, just like in the case of the 3D free electron gas.} \label{f2} 
\end{figure}

Here we derive the density of states (DOS) per unit volume for the energy dispersion Eq.~(\ref{ham3}). The calculation is performed for the 3D case. The same procedure is applied for the 2D case later on (Sec.~\ref{sec2D}).
By definition, the DOS per unit volume is
\begin{equation}\label{g1}
  N(\e) = \frac{2}{V} \sum_\kk\de(\e-\e_{\kk}).
\end{equation}
We use the dimensionless variables defined in the previous section for carrying out the calculation. We change the sum in Eq.~(\ref{g1})  into  integral in spherical
coordinates 
\begin{equation}\label{g2}
 N(\om) =\frac{1}{\pi^2} \sqrt{\frac{A}{B^3}} \int \hspace{-0mm} \ka^2 \, d\ka \, \delta \left(\om - \sqrt{(1- \ka^2)^2 + \D^2 }\right). 
\end{equation}
The delta function in Eq.~(\ref{g2}) is evaluated by decomposing it into a sum 
\begin{equation}\label{g3}
 \de (...) = \sum_{\ka_0} \de(\ka-\ka_0) \, \left| \frac{\om} {2\ka_0(1-\ka_0^2) } \right| ,
\end{equation}
where $\ka_0$ are the positive zeros of the $\de$ function argument
\begin{eqnarray}\label{g4}
\ka_0^{\pm} =  \sqrt{1 \pm \sqrt{\om^2 - \om_b^2}},
\end{eqnarray}
written with the help of  a substitution $\D = \om_b$, defined in Eq.~(\ref{ham5}).
Inserting Eq.~(\ref{g3}) back into  Eq.~(\ref{g2}), we obtain almost the final expression for DOS in 3D
\begin{eqnarray}\label{g5}
&& \hspace{-0mm} N(\om) = \frac{1}{\pi^2} \sqrt{\frac{A}{B^3}} \, \frac{|\om|}{2} \sum_{\ka_0}  \frac{\ka_0 }{ |1 - \ka_0^2|}.
\end{eqnarray}
To make zeros $\ka_0^{\pm}$ real, the sub-root function in Eq.~(\ref{g4}) has to be 
positive. This depends on the value of $\om$. It is easy to check that if $\om_b < \om < \om_t$, then both $\ka_0^+$ and $\ka_0^-$ are real, since the sub-root expression always remains positive. On the other hand, for $\om > \om_t $  only $\ka_0^+$ is real. 
All of these restrictions on the allowed intervals of $\om$ and on the sum over any function of $\ka_0$, can be encoded into $f(\ka_0)$ with the help of the Heaviside step function 
$\Theta(\om)$
\begin{eqnarray}\label{guz}
&& \hspace{0mm}  \sum_{\ka_0} f(\ka_0)  =  \Theta(\om - \om_b)\Theta(\om_t-\om) \left[ f(\ka_0^+) +f(\ka_0^-)  \right] \nonumber \\
&& \hspace{15mm} + \, \Theta(\om - \om_t) f(\ka_0^+).
\end{eqnarray}
Using the recipe Eq.~(\ref{guz}) on Eq.~(\ref{g5}) we obtain the final result for the 3D DOS, which  after some trivial rearrangement of the step functions  is
\begin{eqnarray}\label{g6}
 &&   N(\om)  =    N_0^{(3)}  \frac{|\om|}{\sqrt{\om^2 - \om_b^2}} \Theta(|\om| - \om_b) \bigg[  \left(1 + \sqrt{\om^2 - \om_b^2}\right)^{\frac{1}{2}} \nonumber \\
 &&\hspace{10mm}     + \, \Theta(\om_t - |\om|) 
 \left(1 - \sqrt{\om^2 - \om_b^2}\right)^{\frac{1}{2}} \bigg]. 
\end{eqnarray}
Equation (\ref{g6}) is plotted in Fig.~\ref{f2} for different values of the  parameter $\om_b$ in the units of $N_0^{(3)} = \sqrt{A}/(2\pi^2 B^{3/2}) $.
In the NSSM case $(\om_b=0)$, Eq.~(\ref{g6}) gives the DOS with a dome-like shape between the points $\om  = \pm \om_t= \pm 1$, as seen from definition  (\ref{ham5}).
For a finite value of $\om_b$ we obtain the GSM case where DOS has a square root singularity at $\om_b$. Expanding Eq.~(\ref{g6}), we get 
\begin{eqnarray}\label{g7}
 &&   N(\om)  \approx   N_0^{(3)}  \sqrt{\frac{2\om_b}{{\om - \om_b}}}, \hspace{3mm} \om \gtrapprox \om_b \nonumber \\
 &&    N(\om)  \approx   N_0^{(3)}  \sqrt{\om}, \hspace{3mm} \om \gg \om_b.
\end{eqnarray}
Fig.~\ref{f2} also shows the value of $N(\om_t)$ designated by the red circles. 
This shows how fast $\om_t \to \om_b$ as we increase $\D$. 
While $\om_b$ determines the DOS onset, nothing spectacular happens at $\om_t$.
In the high energy limit we obtain the same DOS as for the 3D free electron gas.

\section{The $T=0$ DC transport quantities in 3D}

There are three concentrations of the charge carriers, for example electrons, usually associated with the DC transport. 
These  are the total concentration $n$, the effective concentration $n_{\al \al}$, and the Hall concentration $n_H$. 
All the three concentrations are functions of the Fermi energy $\om_F$.

In the trivial case of a free electron gas, these three concentrations are identical.
As soon as the dispersion becomes more complex, they begin to deviate from one another. This is because these charge concentrations are each associated with a different transport concept.
The simplest of them is a measure of the total charge added into the system, $n$, and it is temperature-independent.
The second and the third one, $n_{\al\al}$ and $n_H$, are temperature-dependent.
However, in the $T=0$ limit, we can give each of them a simple interpretation.
In that limit, $n_{\al\al}$ becomes the average electron kinetic energy at the Fermi level.
Physically, from the classical Hall experiment, one measures the ratio of transversal voltage and longitudinal current, in the limit of a vanishing magnetic field. This ratio is called the Hall coefficient, $R_H$, and has the dimension of inverse concentration. 
Only in the simplest version---parabolic free electron dispersion---can we trivially connect $R_H$ with the total carrier concentration.
For any more complex band dispersion, $R_H$ can only be calculated through semiclassical Boltzmann transport equations, and the resulting concentration is called the Hall concentration $n_H$.

\subsection{Total charge carrier concentration $n$}

The total carrier concentration 
$n$ is accessible immediately from the DOS. Using the integral representation we have 
\begin{equation}\label{n1}
 n(\om_F) = A  \int_0^{\om_F}  N(\om) \, d\om.
\end{equation}
The total concentration of the intrinsic system can be controlled by chemical doping or through electrostatic doping.
Inserting Eq.~(\ref{g6}) in the above expression we obtain 
\begin{eqnarray}\label{n2}
 && \hspace{0mm}   n(\om_F) =      n_0^{(3)} \,  \Theta(\om_F - \om_b) \times \nonumber \\
 &&\hspace{0mm} \bigg[ \left(1 + \sqrt{\om^2_F - \om_b^2}\right)^{\frac{3}{2}} -   \Theta(\om_t - \om_F) 
 \left(1 - \sqrt{\om_F^2 - \om_b^2}\right)^{\frac{3}{2}} \bigg], \nonumber \\
\end{eqnarray}
where we have introduced a constant for this 3D case
\begin{equation}\label{n3}
 n_0^{(3)}=\frac{1}{3\pi^2} \frac{A}{B}\sqrt{\frac{A}{B}}.
\end{equation}
The total concentration $n$ is plotted in Fig.~\ref{f3} (blue dashed line) for several values of the parameter $\D$. To show some of the properties of $n$ as a function of the Fermi level
$\om_F$, Eg.~(\ref{n2}) is approximated in several interesting limits of $\om_F$:
\begin{eqnarray}\label{n4}
 &&   n(\om_F \approx 0)  \approx  3 n_0^{(3)} \om_F, \hspace{3mm}  \om_b=0 \nonumber \\
  &&   n(\om_F)  \approx  3 n_0^{(3)} \sqrt{2\om_b(\om_F-\om_b)},  \hspace{3mm} \om_F \gtrapprox \om_b \nonumber \\
 &&    n(\om_F)  \approx   n_0^{(3)}  \om_F^{3/2}, \hspace{3mm} \om_F \gg \om_b.
\end{eqnarray}
The concentration which corresponds to the filling of the band up to the top of the inverted parabola is $n_t=n(\om_F = \om_t) = 2^{3/2}n_0^{(3)}$ and is indicated  in the 
Fig.~\ref{f3} by the red circle.

\subsection{Effective charge carrier concentration $n_{\al \al}$}
\label{sec:effective3D}

\begin{figure}[tt]
\includegraphics[width=.48\textwidth]{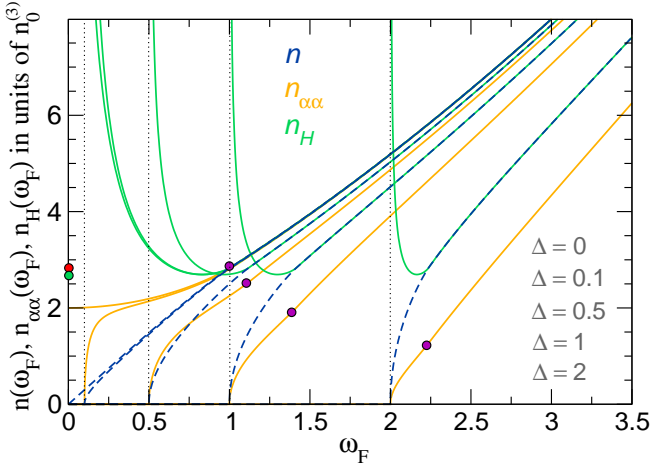} 
\caption{The total concentration $n$ (blue dashed line), the effective concentration $n_{\al \al}$ (orange line) and the Hall concentration $n_H$ (green line) are plotted in units of $n_0^{(3)}$ as functions of Fermi energy $\om_F$,  at several
values of the gap parameter $\D$. The positions of $\D$ are designated by vertical dotted lines.  
It is assumed that $m^* = m_e$. The red circle is the value of $n(\om_t)$, the green circle is the value of $n_H(\om_m)$
while the violet circles represent the values of $n_{\al \al}(\om_t)$.}
\label{f3}
\end{figure}

The effective concentration $n_{\al \al}$  defines the  Drude weight. 
It is a spatially dependent quantity, which at  $T=0$ is given by \cite{Mahan, Rukelj2020}
\begin{equation}\label{n5}
 n_{\al \al}(\ee_F) = \frac{2}{V} \sum_{\kk} m_e v_{\al \kk}  v_{\al \kk} \delta( \e_F -\e_\kk).
\end{equation}
 Here $\al$ is a Cartesian component, $m_e$ is the bare electron mass and  $v_{\al \kk} = (1/\h)\partial \e_\kk /\partial k_\al$ is the electron group velocity.
The derivative of Eq.~(\ref{ham3}) is taken over $\al = x$ (the remaining spatial components give the same result), and inserted in to Eq.~(\ref{n5}). The summation
is changed into integration over the dimensionless variables $\ka$ and $\om$. The evaluation of Eq.~(\ref{n5}) is  similar to the procedure  outlined in the previous subsection.  
The result is 
\begin{eqnarray}\label{n6}
 &&    n_{\al \al}(\om_F) =      n_0^{(3)}\frac{m_e}{m^*} \frac{\sqrt{\om_F^2-\om_b^2}}{\om_F} \, \Theta(\om_F - \om_b) \times  \nonumber \\
 &&   \bigg[ \left(1 + \sqrt{\om_F^2 - \om_b^2}\right)^{\frac{3}{2}}  +  \Theta(\om_t - \om_F) 
 \left(1 - \sqrt{\om_F^2 - \om_b^2}\right)^{\frac{3}{2}} \bigg]. \nonumber \\
\end{eqnarray}
In writing Eq.~(\ref{n6}) we used the definition of the effective mass $m^*$ from the Sec.~\ref{sec1}. 
The resulting expression is relatively similar to Eq.~(\ref{n2}).  Apart from the  ${m_e}/{m^*}$ and the additional prefactor, there is also a sign 
difference within the brackets.
Comparison of  $n_{\al \al}(\om_F)$ and $n(\om_F)$ is shown in Fig.~\ref{f3}.
The differences and similarities between the two carrier concentrations are most noticeable  in the following limits:
\begin{eqnarray}\label{n7}
 &&   n_{\al \al}(\om_F \approx 0)  \approx  2n_0^{(3)}\frac{m_e}{m^*}  \left(1+ \frac{3}{8}\om_F^2 \right), \hspace{3mm}  \om_b=0 \nonumber \\
  &&   n_{\al \al}(\om_F)  \approx  2n_0^{(3)}\frac{m_e}{m^*} \sqrt{2(\om_F-\om_b)/\om_b},  \hspace{3mm} \om_F \gtrapprox \om_b \nonumber \\
 &&    n_{\al \al}(\om_F)  \approx   n_0^{(3)} \frac{m_e}{m^*} \om_F^{3/2}, \hspace{3mm} \om_F \gg \om_b.
\end{eqnarray}
Within the energy range $\om_F \in (0,1)$ the difference between the  $n$  and the $n_{\al \al}(\om_F)$ gives a fingerprint of the NSSM phase. 
The former has a linear--like dependence on the Fermi energy while the later is nearly constant.
The GSM phase is characterized by  $(n, n_{\al \al}) \propto \sqrt{\om_F-\om_b}$  for $\om_F$ just above $\om_b$. 
However, there is a subtle difference between the square root amplitude of the two concentrations. From Fig.~\ref{f3} we see a large square root amplitude in $n_{\al \al}$, while  $n$ is nearly a straight line for small $\D$.
This behavior is  flipped for bigger $\D$.
The value of the effective concentration at the band peak $n_{\al \al}(\om_t)$ is diminishing as we increase $\D$. This is seen by the position of the violet
circles in Fig.~\ref{f3}. In the high energy limit $\om_F \gg \om_t$, the bands (\ref{ham3}) are free-electron-like with the effective mass $m^*$. In this limit we recover a well known 
property $n_{\al \al}/n = m_e/m^*$ \cite{Rukelj2020a}. 

\subsection{Hall concentration $n_H$ and coefficient $R_H$}\label{hall}

Hall concentration $n_H$ of conducting electrons has a fundamental role in the transport equations in the presence of a weak external magnetic field.
$n_H$ changes sign at the energies where the electrons make way for holes as the dominant charge carriers. 
This can be accomplished either by doping \cite{Badoux2016, Zhu2009} or by changing the temperature \cite{Martino2019, LeBoeuf2007}.
The exact derivation of $n_H$ from the semi-classical transport equations is rather complicated and can be found in \cite{Samanta2021, Kupcic2017a, Ong1991, Zener}. 
Here we only recall the limiting low-field value of $n_H$ for a magnetic 
field pointing in $z$ direction and when $T=0$:
\begin{equation}\label{n8}
 n_H = n_{xx}n_{yy}/n_{xy},
\end{equation}
where the concentrations $n_{xx}$ and $n_{yy}$ are given by Eq.~(\ref{n5}). For our specific model, we have $n_{xx}n_{yy} = n^2_{\al \al}$ from Eq.~(\ref{n6}), while $n_{xy}$ is defined as
\begin{equation} \label{n9}
n_{xy } = \frac{2}{V}\sum_{\kk} m_e \left(v_{x \kk}v_{y\kk} M^{yx}_{\kk} - v_{x \kk}v_{x\kk} M^{yy}_{\kk}  \right) \delta(\e_F -\e_\kk).
\end{equation}
$M^{\al \beta}_{\kk}$ is a reciprocal effective mass tensor 
\begin{equation}\label{n10}
 M^{\al \beta}_{\kk} = \frac{m_e}{\h^2} \frac{\partial^2 \e_{\kk}}{ \partial k_\al \partial k_\beta}.
\end{equation}
The Hall coefficient is $R_H = 1/(en_H)$ where $e=-|e|$ is the charge of an electron. 
Taking the first and second derivatives of the dispersion Eq.~(\ref{ham3}) it is straightforward to show, Appendix \ref{2dhall}, that
\begin{equation} \label{n11}
n_{xy } = \frac{16}{V}\frac{m_e^2}{\h^4} B^3\sum_{\kk} k_x^2 \frac{(A - Bk^2)^3}{\e_\kk^3} \delta(\e_F -\e_\kk).
\end{equation}
After a direct evaluation of Eq.~(\ref{n11}) by the procedure outlined in Sec.~\ref{dos} and as shown in Appendix \ref{2dhall}, we obtain an interesting result:
\begin{equation} \label{n11a}
n_{xy}(\om_F) = n(\om_F) \frac{m_e^2}{{m^*}^2} \frac{\om_F^2 - \om_b^2}{\om_F^2}.
\end{equation}
Now we have all the ingredients to calculate the Hall concentration $n_H(\om_F)$ from Eq.~(\ref{n8}): 
\begin{eqnarray}\label{n12}
 &&\hspace{-3mm} n_H(\om_F)/n^{(3)}_0  = \Theta(\om_F - \om_b)\Theta(\om_t - \om_F)  \times \nonumber \\
 &&  \frac{ \left[ \left( 1 + \sqrt{\om_F^2 - \om_b^2}\right)^{\frac{3}{2}} +  \left( 1 - \sqrt{\om_F^2 - \om_b^2}\right)^{\frac{3}{2}} \right]^2 }{ \left(1 + \sqrt{\om_F^2 - \om_b^2}\right)^{\frac{3}{2}} -  \left( 1 - \sqrt{\om_F^2 - \om_b^2}\right)^{\frac{3}{2}}} \nonumber \\
  &&\hspace{5mm}   +  \left( 1 +  \sqrt{\om_F^2 - \om_b^2}\right)^{\frac{3}{2}} \Theta(\om_F -\om_t). 
\end{eqnarray}
$n_H$ as a function of Fermi energy $\om_F$ is shown in  Fig.~\ref{f3} (green line) for several values of the parameter $\D$.
To clarify the main features of $n_H$  we expand Eq.~(\ref{n12}) for specific  limits of $\om_F$ as we did in the case of $n$ and $n_{\al \al}$:
\begin{eqnarray}\label{n13}
 &&   n_H(\om_F \approx 0)  \approx  n_0^{(3)}\frac{4}{3 \om_F}, \hspace{3mm}  \om_b=0 \nonumber \\
  &&   n_H(\om_F)  \approx  n_0^{(3)}\frac{4}{3\sqrt{2\om_b(\om_F-\om_b)}},  \hspace{3mm} \om_F \gtrapprox \om_b \nonumber \\
 &&    n_H(\om_F)  =   n(\om_F), \hspace{3mm} \om_F \geq \om_t.
\end{eqnarray}
As seen from Fig.~\ref{f3}, unlike $n$ and $n_{\al \al}$,  $n_H$ is not a monotonic function of $\om_F$. 
It diverges in both NSSM and GSM phases when $\om_F$ is zero or just above the bottom energy $\om_b$. 
Then as $\om_F$ is increased it drops to a minimal value only to continue to grow again.
The energy $\om_F= \om_{m}$ which corresponds to the minimum of $n_H$ is determined by equating the first derivative of Eq.~(\ref{n12}) to zero.
This can be done analytically and it gives $\om_m = \sqrt{\om^2_b + \eta}$, where $\eta = (3\sqrt{17}-11)/2 \approx 0.685$. 
Correspondingly $n_H(\om_m) \approx 2.694 n_0^{(3)}$, a value designated by the green circle on the $y$ axes in Fig.~\ref{f3}.
Clearly, the energy range in which $n_H$ decreases and exhibits a minimum is within the range $[\om_b, \om_t]$ and is associated with the inverted part of the conduction band. 
Above $\om_t$, $n_H$ is equal to the total concentration $n$ and increases with $\om_F$.

\begin{figure}[tt]
\includegraphics[width=.455\textwidth]{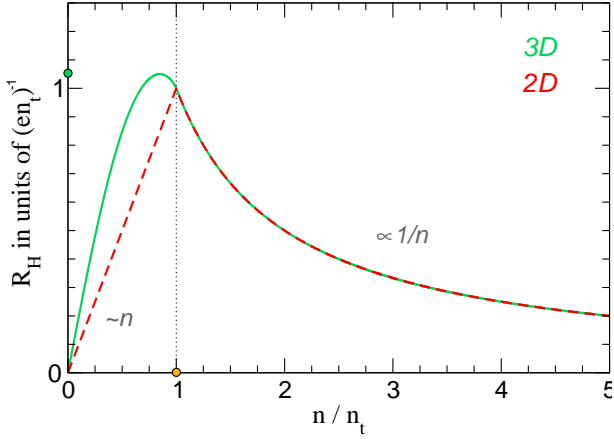} 
\caption{The Hall coefficient $R_H$ for a 3D (green line) and 2D (red dashed line) system as a function of concentration $n$.
The concentration is given in units of concentration needed to fill up the dome of the inverted band $n_t$.
In 3D $n_t= 2^{3/2}n^{(3)}_0$ while in 2D $n_t = 2n_0$.
Correspondingly,  $R_H$ is given in units of $(e n_t)^{-1}$. The two regions  $n < n_t$ and $n > n_t$ 
are divided by a vertical dotted line. Bellow the critical ratio of $1$ (orange circle) for both 3D and 2D $R_H \sim n$, while above
$R_H \propto 1/n$.}
\label{f33}
\end{figure}

Usually the Hall coefficient $R_H$ is expressed as a function of total concentration $n$.
To do so, we invert the Eq.~(\ref{n2}) and find $\om_F$ as function of $n$. We then insert $\om_F(n)$ into Eq.~(\ref{n12}). 
For the 3D case, this procedure is done numerically, yielding $n_H(n)$ and consequently $R_H = 1/(en_H(n))$ shown in Fig.~\ref{f33} in green.
$R_H$ is plotted as a function of scaled concentration $n/n_t$ where, as noted before $n_t=n(\om_t)= 2^{3/2}n_0^{(3)}$. 
If the concentration is small enough, then we have $R_H \sim n$. Above $n/n_t = 1$, $R_H \propto 1/n$. The maximum of $R_H$ is located at the same point as the minimum 
of $n_H$. In scaled units this point is located at  $n/n_t = n(\om_m)/n_t  \approx 0.848$, and gives the height of the peak  $R_H \approx 1.05\, (en_t)^{-1}$
as indicated by the green circle in Fig.~\ref{f33}.
This shows again that Hall coefficient is a function of total electron concentration, above the critical doping $n_t$, just as we expect it should be for a free electron gas.
However, below this critical concentration, it has an unexpected linear-like decrease.

Finally, Eq.~(\ref{n12}) applies only at $T=0$. For finite temperatures the Fermi-Dirac distribution derivative replaces the delta function.
This alters the end result, because of the temperature dependence of the electron chemical potential. 

The results in this Section clearly show that in this system all three carrier concentrations are mutually different. 
The intrinsic NSSM case is characterised  by a zero value of $n$, a constant value of $n_{\al \al}$ and a diverging $n_H$. The intrinsic GSM phase, on the on the other hand,
has a zero value of $n$ and $n_{\al \al}$, while $n_H$ diverges again. 
This difference between the concentrations is even more evident once we step down a dimension, moving from 3D to 2D.

\section{Optical conductivity}\label{optika}

\subsection{General optical conductivity formula}

In a two-band model we can write the complex interband conductivity tensor \cite{Kupcic2013, Rukelj2020}
\begin{equation}\label{op01}
 \s_{\al}(\E) = \frac{2i\hbar}{V}  \sum_{s \neq s'=(c,v)} \sum_{\kk} \frac{|J^{ss'}_{\al \kk}|^2}{\ee^s_{\kk}-\ee^{s'}_{\kk}} \frac{f^{s'}_\kk - f^s_\kk}{\E-\ee^s_{\kk}+\ee^{s'}_{\kk} +i\Gamma}, 
\end{equation}
where  $\Gamma$ is a phenomenological interband relaxation rate, and the Cartesian component $\al$ of the  interband conductivity tensor  is a function of the incident photon energy $\E$.
In Eq.~(\ref{op01}) the  $\al$-dependent interband current vertices
$J^{vc}_{\al \kk} $ are calculated from the Hamiltonian (\ref{ham1}) in Appendix \ref{apa}.
We analytically evaluate the real part of the conductivity tensor (\ref{op01}) in the $\Gamma \to 0$ limit  for the two band model Eq.~(\ref{ham1}).
The result is 
\begin{eqnarray}\label{op1}
&& {\rm{Re}} \,  \s_{\al}(\E) =  \frac{2 \hbar\pi}{V} \sum_{\kk} |J^{vc}_{\al \kk}|^2 \frac{f^v_\kk - f^c_\kk}{\ee^c_\kk-\ee^v_\kk}\delta(\E -\ee^c_\kk + \ee^v_\kk ). \nonumber  \\
\end{eqnarray}
The Fermi-Dirac distributions in the above expression are simplified by taking into account the symmetry of
the bands $\ee^c_\kk = -\ee^v_\kk$ and the fact that the expression Eq.~(\ref{op1}) is finite only for
$\E = \ee^c_\kk -\ee^v_\kk = 2\ee^c_\kk $.  We define
\begin{equation}\label{op2}
\hspace{-0mm}\mathscr{F}(\E) = f^v_\kk - f^c_\kk =  \frac{{\rm{sinh}}(\beta \E/2)}{{\rm{cosh}}(\beta \mu) + {\rm{cosh}}(\beta \E/2)}.
\end{equation}
In the $T = 0$ case, the above expression simplifies to $\Theta(\E -2\ee_F)$, which describes
the suppression of the interband transitions due to the Pauli blocking.
Finally we arrive at a simple expression for the real part of the optical conductivity 
\begin{equation}\label{op3}
{\rm{Re}} \,  \s_{\al}(\mathscr{E}) =   \frac{\mathscr{F}(\E)}{\E} \frac{2 \hbar\pi}{V} \sum_{\kk} |J^{vc}_{\al \kk}|^2 \delta(\E -2\ee^c_\kk ). 
\end{equation}

\subsection{Optical conductivity of the 3D case}

 \begin{figure}[tt]
\includegraphics[width=.46\textwidth]{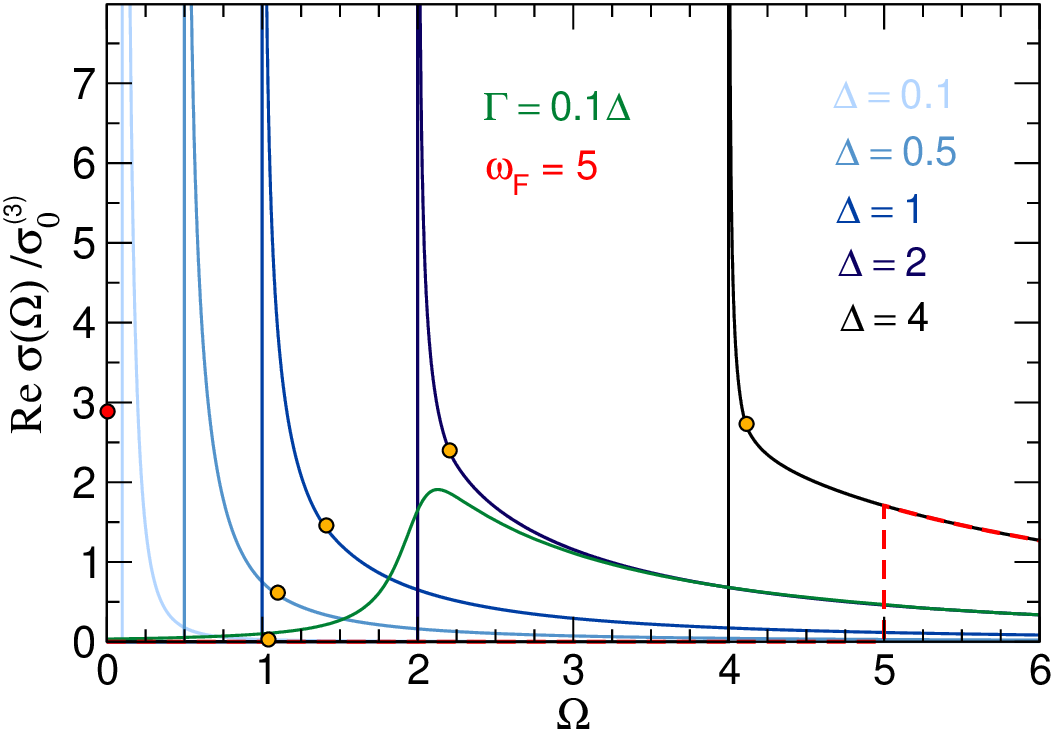} 
\caption{{The real part of the optical conductivity calculated from the two band model (\ref{ham1}) in units of $\s_0^{(3)}$.
${\rm{Re}}\,  \s(\Omega)$ 
is plotted for several values of the parameter $\D$ as a function of a dimensionless parameter $\Om = \mathscr{E}/(2A)$.
The solid lines depict the intrinsic case $(\om_F =0 \to  \mathscr{F}(\Om)=1)$ while the dashed red line gives 
a doped case with $\om_F = 5 \to \mathscr{F}(\Om) = \Theta(\Om-\om_F)$.
All the curves represent the situation of zero interband relaxation as given by Eq.~(\ref{s3}).  
For the specific case of $\D=2$, a finite interband relaxation in Eq.~(\ref{op01}) was used (green line). Orange circles represent ${\rm{Re}}\,  \s(\Omega_t)$. }}
\label{f4}
\end{figure}

We now calculate the real part of the interband conductivity 
 ${\rm{Re}} \, \s_{\al}(\E)$ defined from expression Eq.~(\ref{op3}). The 
 interband current vertex is derived in Appendix \ref{apa}  and it is 
\begin{equation}\label{s1}
 J_{\al \kk}^{vc} =  2 \frac{e}{\h} B C \frac{  k_\al }{\e_\kk}. 
\end{equation}
In our model, $J_{\al \kk}^{vc}$ is a real quantity proportional to the band gap parameter. By inspecting the Hamiltonian (\ref{ham1}) we see that by setting the off-diagonal 
elements ($C$) to zero, there is nothing that could induce the transition between the diagonal elements. Here  this is demonstrated by the shape of $J_{x \kk}^{vc}$ which states the same thing. We conclude that in the NSSM phase there are no optical excitations.

In the GSM phase, we insert the current vertex Eq.~(\ref{s1}) into Eq.~(\ref{op3}) and change the summation to the integral as we did in the Sec.~\ref{dos}.
Once again, we make use of the dimensionless variables $\ka$ and $\Om$ which we now define as $\Om = \E/(2A)$. With this choice of scaling, $\Om$ and $\om$ become equal.
This is best seen by looking at the delta function argument within the integral form of the conductivity in Eq.~(\ref{op3}) 
\begin{eqnarray}\label{s2}
 &&\hspace{-10mm}  {\rm{Re}} \,  \s_\al(\Om) =  \frac{4\s_0}{3\pi}   \sqrt{\frac{A}{ B}}   \mathscr{F}(\Om)   \frac{\Om_b^2}{\Om^3} \times \nonumber \\
 &&\hspace{5mm}  \int \hspace{-0mm} \ka^4  d\ka \,  \delta \left(\Om - \sqrt{(1- \ka^2)^2 +  \D^2}\right). 
\end{eqnarray}
The aforementioned substitutions  enable us to retain the energy scales $\om_b$ and $\om_t$, which we had defined in the Sec.~\ref{sec1}, and which are 
for this purpose renamed to $\Om_b$ and $\Om_t$.
The conductivity constant has been defined in the previous expression $\s_0 = e^2/(4\h)$. Omitting the index $\al$ in the conductivity from now on, the $\delta$ function in Eq.~(\ref{s2}) is solved by a decomposition into a sum of zeros Eq.~(\ref{g4}). Using the recipe (\ref{guz}) we get
\begin{eqnarray}\label{s3}
 &&  {\rm{Re}} \,  \s(\Om) = \s_0^{(3)}   \frac{\Om_b^2}{\Om^2} \frac{\mathscr{F}(\Om)}{ \sqrt{\Om^2 - \Om_b^2} }   \Theta(\Om - \Om_b) \times \nonumber \\
 && \bigg[  \left(1 - \sqrt{\Om^2 - \Om_b^2}\right)^{\frac{3}{2}}  + \Theta(\Om_t - \Om) 
 \left(1 + \sqrt{\Om^2 - \Om_b^2}\right)^{\frac{3}{2}} \bigg] \nonumber \\
\end{eqnarray}
Conductivity Eq.~(\ref{s3}) is shown for the GSM case in Fig.~\ref{f4} for several values of the gap parameter $\D$ in units of $\s_0^{(3)} =  (2\s_0/3\pi)   \sqrt{{A}/{ B}}$. From the expression  Eq.~(\ref{s3}) we immediately see that the amplitude of ${\rm{Re}} \,  \s(\Om)$ depends on the gap value $\Om_b$, which implies that the optical conductivity vanishes in the NSSM case as stated earlier. However, the most striking feature is the divergence of ${\rm{Re}} \,  \s(\Om)$ at the energy $\Om_b$. This divergence is of the square-root type as it can be seen from the expansion of Eq.~(\ref{s3}) for $\Om$ just above $\Om_b$:
\begin{eqnarray}\label{s4}
  &&  {\rm{Re}} \,  \s(\Om) \approx \frac{ 2\s_0^{(3)}}{\sqrt{2\Om_b(\Om-\Om_b)}},  \hspace{3mm} \Om \gtrapprox \Om_b \nonumber \\
 &&   {\rm{Re}} \,  \s(\Om)  \approx   \s_0^{(3)} \Om_b^2 \Om^{-3/2}, \hspace{3mm} \Om \gg \Om_b.
\end{eqnarray}
The divergence in the optical response for the energy $\Om_b$ is seen only in the intrinsic case, when the Fermi energy is zero and $\mathscr{F}(\Om) = 1$. As soon as the doping becomes finite,
the Pauli blockade removes this divergence from the interval of accessible excitation energies (red dashed line in Fig.~\ref{f4}). 
In the intrinsic case of a simple 3D Schr\"odinger-like direct-gap insulator, the onset of the optical transitions is connected with a transition between two points \cite{Cardona}. These two points are the top of the valence band and the  bottom of the conduction band. 
The onset of the optical transitions of the GSM case is characterized by the excitations of the entire surface of points from $-\om_b$ to $\om_b$. This leads to the divergent response  in the optical conductivity.
The ``amplitude'' of the divergence in the real part of the conductivity is also governed by $\Om_b$. The lower this energy is, the more profound the singularity, as seen from
Eq.~(\ref{s4}) and as shown in Fig.~\ref{f4} from comparing the blue curves of different shades.

A finite doping removes the divergence in the optical spectrum. Similarly, this divergence is lifted by taking a finite interband relaxation $\Gamma$ in the calculation of ${\rm{Re}} \,  \s(\Om) $. This leads to the removal of the singularity since at the $\Om_b$ we have 
${\rm{Re}} \,  \s(\Om_b) \propto 1/\Gamma $. For even larger values of $\Gamma$, the optical response of the intrinsic case is entirely smeared, as seen in Fig.~\ref{f4}.

On the other hand, the high $\Om$ limit is proportional to $\Om_b^2$ and  dies off quickly as we lower the value of $\Om_b$, see Fig~\ref{f4}.
The upper dome energy $\Om_t$ bears no significance in  ${\rm{Re}} \,  \s(\Om_t)$. Its position is depicted by the orange circles in Fig.~\ref{f4}. The limiting value of these orange circles 
 approaches $\s_0^{(3)}2^{3/2}$ (shown by the red circle) when $\D$  increases. The distance between $\Om_b$ and $\Om_t$ determines the width of the conductivity peak which is located between these two points. From the definition of  $\om_b$ and $\om_t$ in Eq.~(\ref{ham5}), it is evident that the width decreases with increasing $\D$.
 A final note about Eq.~(\ref{s3}): the optical conductivity as a function of the photon energy $\E$ and the three parameters of the Hamiltonian
 (\ref{ham1}) $A,B$ and $C$ can be easily obtained. In Eq.~(\ref{s3}) we simply need to change $\Om \to \E/(2A)$ and $\Om_b \to C/A$.

\section{2D case}\label{sec2D} 

\subsection{ DOS and the charge  concentrations}
  \begin{figure}[tt]
\includegraphics[width=.45\textwidth]{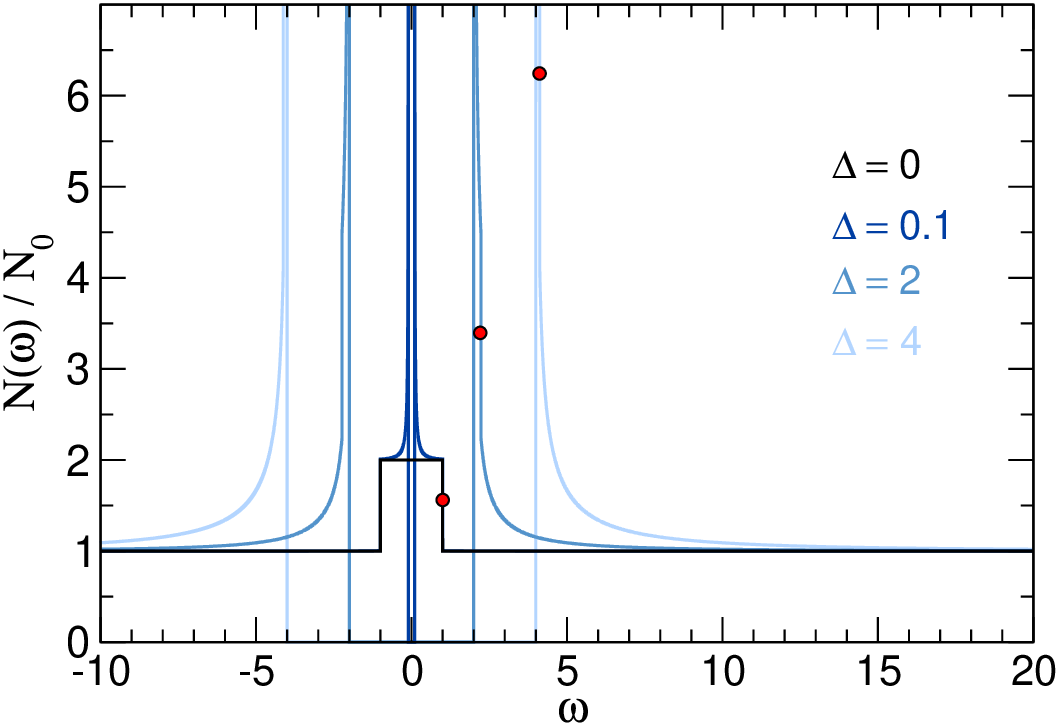} Eq.~(\ref{s2})
\caption{The DOS of the 2D system described by the energy dispersion  $\ee^{c,v}_{\kk} = \pm  \sqrt{( A-  B k^2)^2 +  C^2  }$ [Eq.~(\ref{ham3})] as a function of $\om$, plotted for several values of the parameter $\D$.
For the metallic NSSM case $(\D=0)$, DOS has a step-like shape (black line), while for the gapped GSM case $(\D > 0)$ the divergences appear at $\om_b = \D$. The step feature remains visible for 
$\om_b < 1$. The red circles indicate $N(\om_t)$. In the high $\om$ limit, DOS becomes constant.} \label{f5} 
\end{figure}

So far, we have described the single-particle transport and optical properties of the 3D GSM and NSSM phases. In this section we repeat a similar analysis for
the two dimensional version of the system described in Sec.~\ref{sec1}. Hence, the main difference is the dimension of the integral which needs to be evaluated for various quantities.
We start with writing the end result for  DOS 
\begin{equation}\label{2n1}
N(\om)  =    N_0\frac{|\om|}{\sqrt{\om^2 - \om_b^2}} \Theta(|\om| - \om_b) \Big[ 1 +   \Theta(\om_t - |\om|) \Big],
\end{equation}
where a helpful variable $N_0 = 1/(2\pi B)$ has been introduced.
The DOS in Eq.~(\ref{2n1}) is shown in Fig.~\ref{f5}. The differences are apparent, when compared to its 3D analog in Fig.~\ref{f2}. For the NSSM case a round dome is replaced by a  
step-like structure spanning between $\om \in (-1,1)$. It has an amplitude of $2 N_0$ within the $\om \in (-1,1)$ interval and the height of $ N_0$ outside these boundaries. 
The finite gap in the GSM phase,
like in its 3D analog, introduces a square-root divergence at $\om_b$ \cite{Nicol200}. The DOS for the NSSM case is consistent with the result obtained in \cite{Barati2017}.

\begin{figure}[tt]
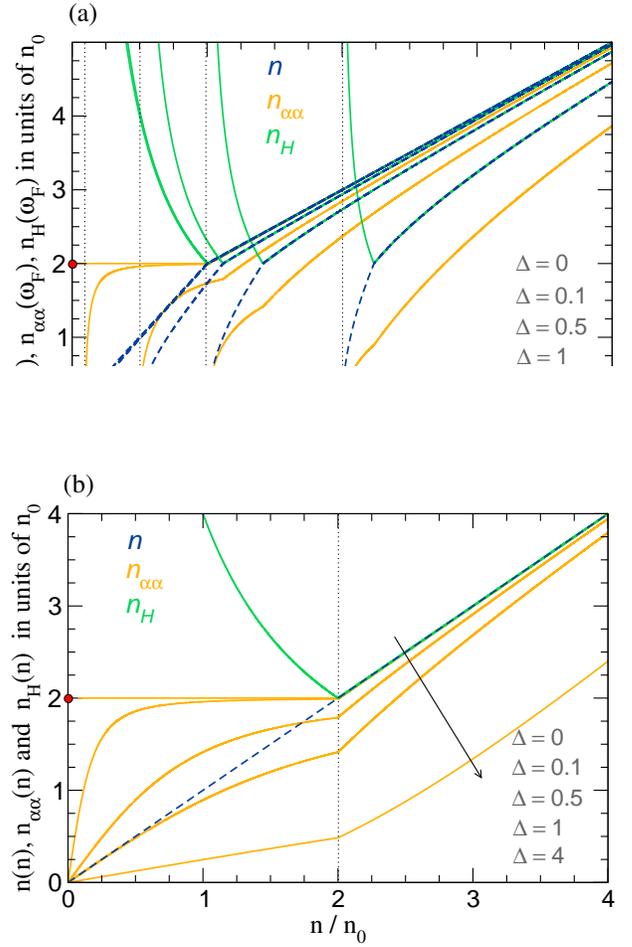

\includegraphics[width=.45\textwidth]{n-o-n.eps} 
\includegraphics[width=.45\textwidth]{2d-novi.eps} 
\vspace*{0mm}
\caption{(a) The total concentrations $n$ Eq.~(\ref{2n2}) (blue dashed line), effective concentration $n_{\al \al}$ Eq.~(\ref{2n4})  (orange line), and the Hall concentration $n_H$ Eq.~(\ref{2n5})
  (green line), as a function of Fermi energy $\om_F$ in units of $n_0$ for the case $m_e = m^*$. The concentrations are plotted for several values of the $\om_b = \D$ whose positions
  are denoted by the vertical dotted lines.
Particularly interesting is the NSSM case $(\om_b=0)$, where the difference between the concentrations is the most profound. All three concentrations have a kink at the $\om_t$. For $n$ and $n_H$ the value at $\om_t$ is $2n_0$ as indicated by the red dot. 
(b) $n$ (blue dashed line), $n_{\al \al}$ from Eq.~(\ref{2n7}) (orange line) and $n_H$  from Eq.~(\ref{2n6}) (green line) are shown as a function of $n$ for several values of $\D$, for the case when $m_e = m^*$.
The arrow indicates the direction of increasing $\D$ in $n_{\al \al}$.
$n$ and $n_H$ do not depend explicitly on $\om_b$ while $n_{\al \al}$ does. Vertical dotted line designates $2n_0$, which is the concentration needed to fill the system to the top of the dome in Fig.~\ref{f1}(b).} \label{f6} 
\end{figure}

The total concentration $n$ is given as a function of the Fermi energy for the electron-doped case $(\om_F > 0)$. It follows from Eq.~(\ref{n1}) with DOS given by Eq.~(\ref{2n1}):
\begin{eqnarray}\label{2n2}
 &&   n(\om_F)  =   n_0\,  \Theta(\om_F - \om_b) \times  \nonumber \\
 &&  \hspace{0mm}  \bigg[ 1 + \sqrt{\om_F^2 - \om_b^2}  -  \Theta(\om_t - \om_F) \left(1 - \sqrt{\om_F^2 - \om_b^2} \right) \bigg]. \nonumber \\
\end{eqnarray}
Once again we introduced a useful constant for 2D concentration $ n_0= A/(2\pi B)$.
The total  concentration $n$ is shown in Fig.~\ref{f6}a as a dashed blue line. In the NSSM case as inherited from the DOS, $n \propto \om_F$ below and above $\om_F =1$
but with different slopes. At this specific energy, which in the GSM case corresponds to $\om_t$, the concentration has a value of $n_t=n(\om_t)=2n_0$ and a kink  in its first derivative over $\om$. 

The effective concentration follows in the same way as in Sec.~\ref{sec:effective3D}. After inserting the electron velocities in the Eq.~(\ref{n5}) and changing it into a 2D integral, we obtain
\begin{eqnarray}\label{2n4}
 &&   n_{\al \al}(\om_F)  =    n_0\frac{m_e}{m^*} \frac{\sqrt{\om_F^2 - \om_b^2}}{\om_F} \Theta(\om_F - \om_b) \times \nonumber \\
 &&  \hspace{0mm}   \bigg[ 1 + \sqrt{\om_F^2 - \om_b^2}  + \Theta(\om_t - \om_F) \left(1 - \sqrt{\om_F^2 - \om_b^2} \right)  \bigg]. \nonumber \\
\end{eqnarray}
In the NSSM case, $n_{\al \al}(\om_F)$ has a constant value up to $\om_t$ and a linear dependence above $\om_t$, as seen from  Fig.~\ref{f6}a where it is depicted with an orange line. The discrepancies between $n$ and $n_{\al \al}$ remain visible over the whole range of values of gap parameter $\D$. For small values of $\om_F$
just above the $\om_b$, $n_{\al \al}(\om_F)$ has a square-root dependence on $\om_F$, while in the high energy limit it goes linearly with $\om_F$.

The final concentration to consider is Hall concentration $n_H$. Since there is a small subtlety in the derivation procedure, we detail it in the Appendix \ref{2dhall} using the recipe from Sec.~\ref{hall}.
It leads to:
\begin{eqnarray}\label{2n5}
 && \hspace{-10mm}   n_H(\om_F)  =   n_0   \bigg[ \frac{2}{\sqrt{\om_F^2 - \om_b^2}} \, \Theta(\om_F - \om_b)\Theta(\om_t-\om_F ) \nonumber \\
 &&  \hspace{5mm}  + \,  \left(1 + \sqrt{\om_F^2 - \om_b^2} \right) \Theta(\om_F - \om_t) \bigg]. 
\end{eqnarray}
Here $n_H$ is drawn as a green line in Fig.~\ref{f6}a, and it has a $\propto 1/n$ dependence below $\om_t$. Equivalently, it has a square-root type of divergence as a function of $\om_F$ when $\om_F \approx \om_b$ (dotted vertical lines). On the other hand, for $\om > \om_t$, it is equal to $n_H = n$. 
Furthermore, inverting  Eq.~(\ref{2n2}) to get $\om_F(n)$ and inserting it in Eq.~(\ref{2n5}) we derive the Hall coefficient as a function of $n$:
\begin{equation}\label{2n6}
n_H(n)  =     \frac{(2n_0)^2}{n} \Theta(2n_0 - n) + n \, \Theta(n -2n_0).
\end{equation}
 $2n_0$ is the total concentration of electrons when the conduction band is filled to the top of the dome of the inverted band ($n_t = 2n_0$).
Although not in a simple fashion like (\ref{2n6}), by the same procedure $n_{\al \al}$  too can be written as a function of $n$:
\begin{eqnarray}\label{2n7}
 && \hspace{-10mm}  n_{\al \al}(n)  =   \frac{m_e}{m^*}  \bigg[ \frac{n}{\sqrt{(n/(2n_0))^2 + \om_b^2}} \, \Theta(2n_0 - n) \nonumber \\
 &&  \hspace{5mm}  + \, \frac{n\left(n/n_0 -1 \right) }{\sqrt{(n/n_0 - 1)^2 + \om_b^2}} \, \Theta(n -2n_0) 
  \bigg] .
\end{eqnarray}
The dependence of $n_H$ and  $n_{\al \al}$ on  $n$ is depicted in Fig.~\ref{f6}b. The three concentrations of the 2D system differ from one another for  energies $\om_F < \om_t$, or for total concentration  $n < 2n_0$. 
The differentiation between the two ground states NSSM and GSM phase now becomes easy to make.
By carefully changing the doping and reading out the Drude weight ($ n_{\al \al}$), one should obtain a constant in the NSSM phase for $n < 2n_0$. 
Another valid fingerprint of the band structure (\ref{ham1}) in the 2D transport is the $n_H$, which diverges as $1/n$ for $n < 2n_0$. 
Unlike in the 3D case, the minimum value of $n_H$ is now located at the energy $\om_t$.

In the highly doped limit, where the Fermi energy overshoots the top of the dome $\om_t$, or equivalently where $n > 2n_0$, we expect all the three concentrations to be roughly the same,
$n_H \approx n_{\al \al} \approx n$. This is to be expected since the energy dispersion (\ref{ham3}) high above the dome is free-electron-like. From Eq.~(\ref{2n6}) we see this is true
for $n$ and $n_H$ for GSM and NSSM cases. But in this doping regime,  $n_{\al \al} = n$ only in the NSSM phase, while in the GSM phase it approaches  $n$ asymptotically as the doping increases.

The Hall coefficient as a function of $n$ is reciprocal to Eq.~(\ref{2n6}): 
\begin{equation}\label{2n8}
R_H  =     \frac{n}{n_t^2 e} \Theta(n_t - n) + \frac{1}{ne} \, \Theta(n -n_t),
\end{equation}
and it is shown in Fig.~\ref{f33} (red dashed line) as a function of $n/n_t$, in units of $(e n_t)^{-1}$. 
As in the 3D case, $R_H$ grows linearly with $n$ until it reaches a sharp maximum at $n_t$ with a height of $R_H =1$. Beyond this maximum, $R_H$ has the same $1/n$ dependence like its 3D analog.
Previously, we could find the 3D $R_H(n)$  only numerically.
Fortunately, in the 2D case, we can obtain an analytical result for $R_H$,.
This is why the two-dimensional case provides a valuable insight into the 
signature of the inverted bands in the  $T=0$ magneto-transport \cite{Zhu2009}.

\subsection{Optical conductivity of the 2D system}

\begin{figure}[tt]
\includegraphics[width=.45\textwidth]{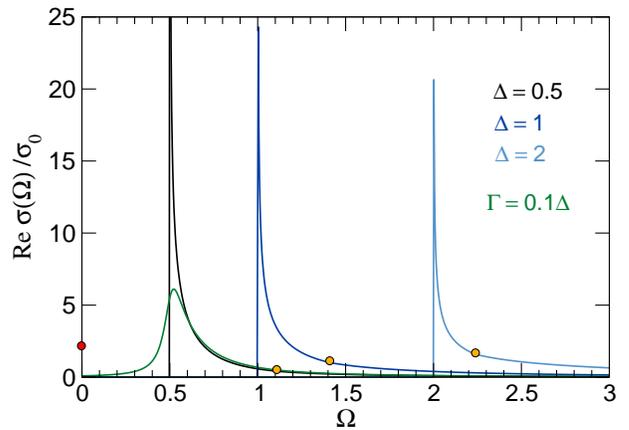} 
\caption{Real part of the optical conductivity of the 2D system described by the two band model (\ref{ham1}) in units of $\s_0$.
${\rm{Re}}\,  \s(\Omega)$ 
is plotted for several values of parameter $\D$ as a function of  $\Om = \mathscr{E}/(2A) $ for the intristic case $(\mathscr{F}(\Om)=1)$.
The green curve (with a finite $\Gamma$) is calculated using the Eq.~(\ref{op01}) with $\Gamma = 0.05$~eV, while the rest is given by Eq.~(\ref{2ds2}). The step-like feature in the DOS is not visible
in the conductivity.  ${\rm{Re}}\,  \s(\Omega_t)$ is depicted by orange circles. The circles approach $2\s_0$ as $\D \to \infty$ (red circle).} \label{f7} 
\end{figure}

We obtain the integral expression for the 2D GSM optical conductivity
by inserting the interband current vertex Eq.~(\ref{s1}) in the real part of the interband conductivity formula Eq.~(\ref{op3}).
In units of $\Om$ and dimensionless $\ka$, it reads
\begin{eqnarray}\label{2ds1}
 {\rm{Re}} \,  \s(\Om) =  &&\frac{e^2}{2\hbar}  \frac{\mathscr{F}(\Om)}{\Om  }  \frac{\Om_b^2}{\Om^2}    \int \hspace{-1mm} \ka^3  d\ka \,  \delta \left(\Om - \sqrt{(1- \ka^2)^2 +  \D^2}\right).  \nonumber \\
\end{eqnarray}
Evaluating the above integral in the same way as in Sec.~\ref{sec2}, we get 
\begin{eqnarray}\label{2ds2}
 &&  {\rm{Re}} \,  \s(\Om) = \s_0   \frac{\Om_b^2}{\Om^2} \frac{\mathscr{F}(\Om)}{ \sqrt{\Om^2 - \Om_b^2} }   \Theta(\Om - \Om_b) \times \nonumber \\
 && \bigg[   1 + \sqrt{\Om^2 - \Om_b^2}  \,  + \,  \Theta(\Om_t - \Om) 
 \left(1 - \sqrt{\Om^2 - \Om_b^2}\right) \bigg] \nonumber \\
\end{eqnarray}
As in the 3D case, we can make use of the conductivity constant $\s_0$. For a simple expression like (\ref{2ds2}), two limiting regimes in $\Om$ are easily found:
\begin{eqnarray}\label{2ds3}
  &&  {\rm{Re}} \,  \s(\Om) \approx \frac{ 2\s_0}{\sqrt{2\Om_b(\Om-\Om_b)}},  \hspace{3mm} \Om \gtrapprox \Om_b \nonumber \\
 &&   {\rm{Re}} \,  \s(\Om)  \approx   \s_0 \Om_b^2 \Om^{-2}, \hspace{3mm} \Om \gg \Om_b.
\end{eqnarray}
The conductivity Eq.~(\ref{2ds2}) is plotted in Fig.~\ref{f7} in units of $\s_0$ for several values of the gap parameter $\D$ in the intrinsic case, $\mathscr{F}(\Om)=1$.
The high energy tail decreases  stronger than in 3D, apart from the root-like divergence near the gap energy, which has the same shape as in the 3D case.
Such a weak response would be difficult  to set apart from the background, if multiple bands are present as they usually are in real systems.
${\rm{Re}} \,  \s(\Om_t)$ is indicated by orange dots and it approaches the red dot $2\s_0$ as $\D \to \infty$.
The impact of the finite interband relaxation $\Gamma$ on the ${\rm{Re}} \,  \s(\Om) $ is depicted in Fig.~\ref{f7} with a green line.

\section{${\rm{Re}} \,  \s(\Om) $ and the JDOS argument}

Usually in a system with linear band dispersion  the shape of the interband conductivity can be be 
determined by the rule-of-thumb argument involving the joint density of state (JDOS) \cite{gruner}. 
To show how this works, we look at the Eq.~(\ref{op3}) in the general two-band case
\begin{equation}\label{jdos1}
{\rm{Re}} \,  \s_{\al}(\mathscr{E}) =   \frac{\mathscr{F}(\E)}{\E} \frac{2 \hbar\pi}{V} \sum_{\kk} |J^{vc}_{\al \kk}|^2 \delta \left(\E -(\ee^c_\kk - \ee^v_\kk)\right). 
\end{equation}
If we assume that the interband current vertex does not depend explicitly on $k$, and that
$J_{\al \kk}^{vc} = J_{\al}^{vc}$, then  $J_{\al}^{vc}$ can be taken outside the sum in Eq.~(\ref{jdos1}). The remaining  sum over the  $\delta$ function  is the definition of JDOS.
If the bands have electron-hole symmetry,  JDOS is equivalent to DOS such that Eq.~(\ref{jdos1}) becomes 
\begin{equation}\label{jdos2}
{\rm{Re}} \,  \s_{\al}(\mathscr{E}) =  \frac{\hbar\pi}{2}   |J^{vc}_{\al}|^2  \frac{\mathscr{F}(\E)}{\E}  N(\E/2)
\end{equation}
where we have used the definition of DOS (\ref{g1}). For the 3D and 2D Dirac systems \cite{AshbyPRB14,Kupcic2015} $J_{\al \kk}^{vc} = ev$ where $v$ is the Dirac velocity and hence (\ref{jdos2}) applies.

For the model studied in this article, the JDOS approach does not apply for small photon energies $\E \sim C$. It is  only in the high-energy limit that the Eq.~(\ref{jdos2}) can be safely applied. The obvious reason for this is the strong $\kk$ dependence of the current vertex (\ref{s1}).
A way around it is to notice that in the high energy limit  ($\e_\kk \gg C$) the dispersion Eq.~(\ref{ham3}) is parabolic, $\e_\kk \approx Bk^2$, and isotropic.
Also in this limit the inverse function $k(\e)$ is single valued, which is not the case when $\e_\kk < \sqrt{A^2 + C^2}$, see Sec.~\ref{sec2}.
Hence in $D$ dimensions the mean square of the component is $\la k_\al^2 \ra = k^2/D = \e_\kk /(DB)$ and the square of the D-dimensional interband current vertex is  
\begin{equation}\label{jdos3}
 |J_{\al \kk}^{vc}|^2 \to \frac{8e^2}{\h^2} \frac{C^2B}{D} \frac{1}{\E},
\end{equation}
where we have used $\E = 2\e_\kk$. Inserting (\ref{jdos3}) into Eq.~(\ref{jdos2}) we get the same expressions, once we change $\E = 2A \Om$, as we did for the high-$\Om$ expansion in 2D Eq.~(\ref{2ds3}) and  3D Eq.~(\ref{s4}).

This line of reasoning for the JDOS rule-of-thumb is not new and can be demonstrated on the example of 
 the massive 2D Dirac system, where, if we designate the band-gap by $C$, we have  $|J_{\al \kk}^{vc}|^2 \propto e^2v^2 (1+C^2/\E^2)$ \cite{Jafari2012}. Since the  DOS of the massive 2D Dirac
 system is linear in $\E$, in this case Eq.~(\ref{jdos2}) gives an exact result for the optical conductivity.

\section{Conclusions}
We have addressed the static and dynamic transport properties of the nodal surface semimetals and their gapped phase.
The properties of these systems in three and two dimensions are described  by the two-band model of the valence electrons.  
The main feature of this model is the inversion of the valence bands below the certain energy and parabolic like shape for energies above it.
The main question we answer is: Can we determine the electronic
ground state, GSM or NSSM, by comparing the experimental and the calculated transport and optical properties?

The band inversion is responsible for a specific shape of the  NSSM density of state. In the 3D case it is a dome-like structure, while in the 2D case it has a step-like feature. In the GSM phase a square-root divergence occurs in the DOS at the band gap energies. 

We have studied  three different concentrations of the charge carriers. These are the total, effective and Hall concentration.
We have shown that  by controlling the doping and 
comparing the three concentrations we can conclude if the ground state is gapped or not. This is due to the fact that both in 3D and in 2D these three concentrations 
differ from one another. Only for high doping (Fermi energy much larger than the band gap) do they become equal. 
The differences between the transport properties are  more profound in 2D than in the 3D case. Still,  the Hall coefficient shows remarkable similarities 
between the 3D and 2D when plotted as a function of total concentration. 

The optical properties give a definitive proof of the ground state. There are no optical excitations in the NSSM phase, since the conductivity amplitude  is proportional to the 
band gap. The optical response of the GSM phase has a square root divergence above the band gap threshold for both 3D and 2D with a  $\s \propto \Om^{-2}$ tail dependence in the 2D and a $\s \propto \Om^{-3/2}$ tail in the 3D case.

Finally, the JDOS rule-of-thumb derivation of the optical conductivity is elaborated in detail. For the GSM phase it is shown to work only in the high-energy limit.

\section{Acknowledgments}
Z.~R. acknowledges the hospitality of the University of Fribourg. 
A.~A. acknowledges funding from the  Swiss National Science Foundation through project PP00P2\_170544.

\appendix

\section{3D  and 2D $n_H$}\label{2dhall}

The electron velocities for $\al = x,y$ components are
\begin{equation}\label{apb1}
 v_{\al \kk} =  -\frac{1}{\h} \frac{2Bk_\al(A-Bk^2)}{\e_\kk}.
\end{equation}
The mass tensor components Eq.~(\ref{n10}) are  
\begin{equation}\label{apb2}
\frac{\h^2}{m_e} M^{yy}_{\kk} =     \frac{4B^2k_y^2}{\e_\kk}- \frac{2B(A-Bk^2)}{\e_\kk} - \frac{4B^2k_y^2(A-Bk^2)^2}{\e^3_\kk}, 
\end{equation}
and
\begin{equation}\label{apb3}
\frac{\h^2}{m_e} M^{yx}_{\kk} =    \frac{4B^2k_x^2k_y^2}{\e_\kk} - \frac{4B^2k_x k_y(A-Bk^2)^2}{\e^3_\kk}. 
\end{equation}
The velocity and mass tensor product within Eq.~(\ref{n9}) is 
\begin{equation}\label{apb4}
   v_{x \kk}v_{y\kk} M^{yx}_{\kk} -  v_{x \kk}v_{x\kk} M^{yy}_{\kk}  =  \frac{m_e}{\h^4} \frac{8B^3k_x^2(A-Bk^2)^3}{\e^3_\kk}. 
\end{equation}
First  the 3D case is solved. (\ref{apb4}) is inserted in Eq.~(\ref{n9}) and  the sum converted to integral with dimensionless variables. 
This integral contains the $\delta$ function which again is decomposed 
as
\begin{equation}\label{apb5}
  n_{xy}(\om_F)  =    \frac{m_e^2}{\h^4} \frac{16}{3(2\pi)^2}\frac{AB}{\om_F^2} \sqrt{\frac{A}{B}} \sum_{\ka_0} \frac{\ka_0^4(1-\ka_0)^3}{|\ka_0(1-\ka_0)|}.
\end{equation}
The zeros $\ka_0$ as defined in Sec.~\ref{sec2} are carefully implemented in (\ref{apb5}) with particular care on the $\pm$ sign of
$1-\ka_0^2 = \pm \sqrt{\om_F^2 - \om_b^2} $. This sign is preserved once taken to the power of 3. We get
\begin{eqnarray}\label{apb6}
 && \hspace{-0mm}   n_{xy}(\om_F) =      n_0^{(3)}  \frac{m_e^2}{{m^*}^2} \frac{\om_F^2-\om^2_b}{\om^2_F} \bigg\{  \Theta(\om_F-\om_b)\Theta{(\om_t -\om_F)} \times \nonumber \\
 &&\hspace{5mm}  \left[ \left(1 + \sqrt{\om^2_F - \om_b^2}\right)^{\frac{3}{2}} - \left(1 - \sqrt{\om^2_F - \om_b^2}\right)^{\frac{3}{2}} \right] \nonumber \\
 &&\hspace{5mm} + \, \,  \Theta(\om_F - \om_t) 
 \left(1 - \sqrt{\om_F^2 - \om_b^2}\right)^{\frac{3}{2}} \bigg\},
\end{eqnarray}
or after we rearrange the $\Theta$ functions 
\begin{eqnarray}\label{apb66}
 && \hspace{-0mm}   n_{xy}(\om_F) =      n_0^{(3)}  \frac{m_e^2}{{m^*}^2} \frac{\om_F^2-\om^2_b}{\om^2_F} \,  \Theta(\om_F-\om_b) \times    \nonumber \\
 &&\hspace{0mm} \left[    \left(1 + \sqrt{\om^2_F - \om_b^2}\right)^{\frac{3}{2}}  -   \Theta(\om_t - \om_F) 
 \left(1 - \sqrt{\om_F^2 - \om_b^2}\right)^{\frac{3}{2}} \right]. \nonumber \\
\end{eqnarray}
Comparing (\ref{apb66}) with Eq.~\ref{n2} we conclude
\begin{eqnarray}\label{apb7}
 && \hspace{-10mm}   n_{xy}(\om_F) =     n(\om_F) \frac{m_e^2}{{m^*}^2} \frac{\om_F^2-\om^2_b}{\om^2_F} 
\end{eqnarray}
In deriving Eq.~(\ref{apb66}) we have used the definition of constant (\ref{n3})
\begin{equation}\label{n3333}
 n_0^{(3)}=\frac{1}{3\pi^2} \frac{A}{B}\sqrt{\frac{A}{B}},
\end{equation}
as well as  $B= \h^2/(2m^*)$.
Finally the Hall concentration follows from the definition $n_H(\om_F) = n_{\al \al}^2(\om_F)/n_{xy}(\om_F)$ as written in main text.

The same procedure applies for the 2D case. 
First we reorganise the $\Theta$ functions within the effective concentration $n_{xx} = n_{yy}= n_{\al \al}$ Eq.~(\ref{2n4})
\begin{eqnarray}\label{apb8}
 &&   n_{\al \al}(\om_F)  =    n_0\frac{m_e}{m^*} \frac{\sqrt{\om_F^2 - \om_b^2}}{\om_F}  \bigg[ 2 \, \Theta(\om_F - \om_b)\Theta(\om_t - \om_F)  \nonumber \\
   &&\hspace{15mm}   +  \left(1 + \sqrt{\om_F^2 - \om_b^2} \right) \Theta(\om_F - \om_t) \bigg]. 
\end{eqnarray}
Next we calculate $n_{xy}$ using Eq.~(\ref{n9}) and Eq.~(\ref{apb4}) 
\begin{eqnarray}\label{apb9}
 &&\hspace{-0mm}  n_{xy}(\om_F) =  \frac{4}{\pi} \frac{m_e^2}{\h^4}    \frac{AB}{\om_F^3} \times \nonumber \\
 &&\hspace{0mm}  \int \hspace{-0mm} \ka^3 (1-\ka^2)^3  d\ka \,  \delta \left(\om_F - \sqrt{(1- \ka^2)^2 +  \D^2}\right) \nonumber \\
  &&\hspace{5mm}   = \frac{4}{\pi} \frac{m_e^2}{\h^4}     \frac{AB}{\om_F^3}  \frac{\om_F}{2}\sum_{\ka_0} \frac{\ka_0^3 (1-\ka_0^2)^3 }{|\ka_0 (1-\ka_0^2) |}. 
\end{eqnarray}
Again, the zeros $\ka_0$ as defined in Sec.~\ref{sec2} are inserted in (\ref{apb9}). We obtain 
\begin{eqnarray}\label{apb10}
 &&\hspace{-10mm}    n_{xy}(\om_F)  =    n_0\frac{m^2_e}{{m^*}^2} \frac{{\om_F^2 - \om_b^2}}{\om^2_F}  \times \nonumber \\
 &&  \hspace{0mm}   \bigg[ 2  \sqrt{\om_F^2 - \om_b^2} \, \Theta(\om_F - \om_b)\Theta(\om_t - \om_F)  \nonumber \\
 && +  \left(1 + \sqrt{\om_F^2 - \om_b^2} \right) \Theta(\om_F - \om_t) \bigg] .
\end{eqnarray}
In writing (\ref{apb10}) we have used $B= \h^2/(2m^*)$ and the definition of $n_0 = A/(2\pi B)$.
Using equations (\ref{apb8}) and (\ref{apb10}) we get 
\begin{eqnarray}\label{apb11}
 && \hspace{-10mm}  n_H(\om_F)  = \frac{n_{\al \al}^2}{n_{xy}}  \nonumber \\
 && \hspace{-5mm} = n_0  \bigg[ \frac{2}{\sqrt{\om_F^2 - \om_b^2}}\Theta(\om_F - \om_b)\Theta(\om_t - \om_F)  \nonumber \\
 && \hspace{5mm} +  \left(1 + \sqrt{\om_F^2 - \om_b^2} \right) \Theta(\om_F - \om_t) \bigg] .
\end{eqnarray}

\section{current vertices}\label{apa}

We start with the general form of the  $2\times 2$ Hamiltonian matrix in non-diagonal representation
\begin{equation}\label{a1}
 \mb{H} =  \begin{pmatrix}
 b_{\kk} & a_{\kk} \\
 a^*_{\kk} & d_{\kk}
\end{pmatrix}. 
\end{equation}
The matrix elements are labeled by $H_{\kk}({\ell, \ell'})$ where $(\ell, \ell')$ are the row and column indices of Eq.~(\ref{a1}). If we label the 
two Bloch energies  by $(s, s') = (c,v)$ we can define  the $\al$ component of the current vertex 
\begin{equation} \label{a2}
  J^{ss'}_{\al\kk} = \sum_{\ell \ell'} \frac{e}{\hbar} \frac{\partial H_{\kk}(\ell, \ell')}{\partial k_{\alpha}}U_{\kk}(\ell,s)U^*_{\kk}(\ell',s'),
\end{equation}
where $U_{\kk}(\ell,s)$ are the elements of unitary matrix. This matrix transforms Hamiltonian to its diagonal form by definition $\mb{U}\mb{H}\mb{U}^{-1} = \mb{E}$, where $\mb{E}$ is the eigenvalue matrix. After a tedious derivation $\mb{U}$ is shown to be 
\begin{equation} \label{a3}
U_{\kk}(\ell, s) =  \begin{pmatrix}
  e^{i\varphi_{\kk}}\cos ({\vartheta_{\kk}}/{2}) & e^{i\varphi_{\kk}}\sin ({\vartheta_{\kk}}/{2})
  \vspace{2mm} \\
  - \sin ({\vartheta_{\kk}}/{2})  &  \cos ({\vartheta_{\kk}}/{2})
 \end{pmatrix},
\end{equation}
where
\begin{equation} \label{a4}
  a_{\kk}  =|a_{\kk}|e^{i\varphi_{\kk}}, \hspace{2mm} \tan \varphi_{\kk} = \frac{{\rm Im}\, a_{\kk}}{ {\rm Re} \, a_{\kk}},\hspace{2mm} \tan \vartheta_{\kk} = \frac{2|a_{\kk}|}{d_{\kk} - b_{\kk}}.
\end{equation}
Therefore  Eq.~(\ref{a1}) and  Eq.~(\ref{a2}) give after some trigonometric manipulation the intraband ($s =s'=c$) current vertex 
\begin{eqnarray}\label{a55}
 && \frac{\hbar}{e}J_{\al \kk}^{cc} = \cos \vartheta_{\kk} \frac{1}{2}  \frac{\partial (b_{\kk} - d_{\kk})}{\partial k_{\alpha}} 
 + \sin \vartheta_{\kk} \frac{\partial |a_{\kk}|}{\partial k_{\alpha}} ,
\end{eqnarray}
and for the interband case ($c=s\neq s'=v$)
\begin{eqnarray}\label{a5}
 && \frac{\hbar}{e}J_{\al \kk}^{vc} = \sin \vartheta_{\kk} \frac{1}{2}  \frac{\partial (b_{\kk} - d_{\kk})}{\partial k_{\alpha}} + i |a_{\kk}|\frac{\partial \varphi_{\kk}}{\partial k_{\alpha}}
 + \cos \vartheta_{\kk} \frac{\partial |a_{\kk}|}{\partial k_{\alpha}}. \nonumber \\
\end{eqnarray}
For the model (\ref{ham1}) $ a_{\kk}  = C$ and $\tan \varphi_{\kk} = 0$ and 
\begin{equation}  \label{a6}
 \frac{ \partial |a_{\kk}|  }{\partial k_{\al}} =  0, \hspace{5mm} \frac{ \partial \varphi_{\kk}  }{\partial k_{\al}}=0.
\end{equation}
The only non vanishing element is the first part on the right hand side of (\ref{a5}) and (\ref{a55}). For the specific case of Hamiltonian   Eq.~(\ref{ham1})
\begin{equation} \label{a8}
 \frac{\partial (b_{\kk} - d_{\kk})}{\partial k_{\alpha}} = -2Bk_\al.
\end{equation}
This in turn gives the final expression for the interband current vertex
\begin{eqnarray}\label{a9}
 J_{\al \kk}^{vc} =   \frac{e}{\h} 2BC  \frac{k_\al}{ \sqrt{( A-  B k^2)^2 +  C^2  }}. 
\end{eqnarray}
In limits $k = 0$  and $k \to \infty$ (\ref{a9}) is 
\begin{equation}\label{a11}
 J_{\al \kk}^{vc} \approx   \frac{e}{\h} 2B k_\al \frac{C}{\sqrt{A^2 + C^2}}, \hspace{4mm}  J_{\al \kk}^{vc} \approx   \frac{e}{\h} 2C \frac{k_\al}{k^2}.
\end{equation}

\bibliography{nssm}

\end{document}